\documentclass[preprint]{revtex4}
\usepackage{graphicx}
\usepackage[latin1]{inputenc}
\usepackage{epsfig}
\usepackage{amsmath}
\usepackage{amsfonts}
\usepackage{amssymb}
\usepackage{perpage} 
\MakePerPage{footnote}

\begin{document}

\title {Higgs-radion unification: radius stabilization by an SU(2) bulk doublet and the 126 GeV scalar}
\author{Michael Geller}
\email{mic.geller@gmail.com}
\affiliation{Physics Department, Technion-Institute of Technology, Haifa 32000, Israel}
\author{Shaouly Bar-Shalom}
\email{shaouly@physics.technion.ac.il}
\affiliation{Physics Department, Technion-Institute of Technology, Haifa 32000, Israel}
\author{Amarjit Soni}
\email{adlersoni@gmail.com}
\affiliation{Theory Group, Brookhaven National Laboratory, Upton, NY 11973, USA}

\begin{abstract}
We investigate a Randall-Sundrum model with an SU(2) doublet propagating in the bulk.
Upon calculating its gravitational effect we find that a stabilized radius can be generated
without the use of an additional scalar, as needed for example in the Goldberger-Wise (GW) mechanism,
and with no additional fine-tuning
other than the inescapable one due to the cosmological constant; similar tuning is also present in
the GW mechanism.
The lowest scalar excitation in this
scenario, the counterpart of the radion of the GW mechanism, has both radion-like and Higgs-like
couplings to the SM fields.
It, thus, plays a dual role and we, therefore, denote it as the ``Higgs-radion" ($h_r$).
As opposed to the GW radion case, our Higgs-radion
is found to be compatible with the 126 GeV scalar recently discovered at the LHC, at the level of $1\sigma$,
with a resulting 95\% CL bound on the KK-gluon mass of:
$4.48~TeV<M_{KKG}< 5.44~TeV$.
An important consequence of our setup should be accentuated: 
the radion of the traditional RS scenarios 
simply does not exist, so that 
our Higgs-radion is not the conventional
mixed state between the GW radion and the Higgs.
\end{abstract}


\maketitle

\section{Introduction}

With the recent discovery of the 126 GeV scalar particle ($0^+$) at the LHC \cite{LHC-Higgs}, it is naturally appealing to identify it with the long sought-for missing piece of the Standard Model (SM), the Higgs boson. However, other attractive interpretations of the 126 GeV Higgs-like state were also considered,
one of them being the radion excitation of the metric in Randall-Sundrum (RS) models of a Warped Extra Dimension \cite{RS} with a dynamically stabilized radius, most notably via the Goldberger-Wise (GW) mechanism \cite{GW,DeWolfe-GW,CSAKI-Radion}. Indeed, in recent years, there has been a growing interest in the radion state due to the fact that it is possibly the lightest new particle
in the RS spectrum, and many aspects of its phenomenology have been thoroughly studied \cite{Radion-phenmenology,CSAKI-Radion,Radion-cosm,CSAKI-REAL,Higgs-Radion-Mixing,Higgs-Radion-Mixing-2,Radion125,Radion-Xsec}

In the GW mechanism \cite{GW}, a coupled system of gravity and a bulk scalar dynamically generate the finite radius of the extra dimension. The Einstein equations of this system can either be solved directly \cite{DeWolfe-GW,CSAKI-Radion} or using an effective 4d approach, as in \cite{GW}, while both are usually solved under the assumption of a small backreaction, which greatly simplifies the calculations. The main difference with pure RS, in which both brane tensions have to be tuned to satisfy the boundary conditions of the Einstein equations, is that the size of the extra dimension explicitly enters the equations. Thus, one of the boundary conditions sets the size of the extra dimension and in the process alleviates one of the fine-tunings in the RS1 scenario. The remaining fine-tuning is that which sets the 4d cosmological constant to zero.
In this scenario, the radion is the zero-mode of both the scalar metric fluctuation and the bulk scalar (in the KK-tower of quantum fluctuations to this vacuum).
Its mass is typically  suppressed (parametrically) with respect to the KK scale and, thus, it is assumed to be lighter than the rest of the KK spectrum.
The radion has the same quantum number as the SM Higgs and it couples to the SM content via the stress-energy tensor \cite{CSAKI-Radion,CSAKI-REAL}.
 It thus has similar decay signatures as the Higgs, albeit with different rates.
 This fact has raised the question of whether the radion interpretation  of the 126 GeV discovery is consistent with the LHC data \cite{Radion125}.
 The answer to that seems to be negative:
 a good fit to the LHC Higgs data yields a KK scale of ${\cal O}(1~TeV)$,
 which is excluded by LHC direct searches \cite{LHC-search}.
 The possibility of a mixed Higgs-radion state is also excluded when this state is radion-dominated \cite{Higgs-Radion-Mixing}.

In this paper we present an alternative radius stabilization scenario in which the bulk scalar has the quantum numbers of a 5d ${SU(2)}_{L}$ bulk-Higgs doublet \cite{bulk-Higgs}.
As we will show, the VEV of this object can both break the EW symmetry and at the same time play the role of the GW scalar in stabilizing the radius of the extra-dimension.  Similar ideas have already appeared in the literature \cite{HiggsRadion,Vecchi}.
In particular, in \cite{Vecchi} a similar setup was investigated
both from the AdS and CFT approaches, using the AdS/CFT correspondence \cite{Holography},$^{[1]}$\footnotetext[1]{As is well known, the radion is dual to a dilaton
arising from the breaking of conformal symmetry at the IR scale, see e.g., \cite{dilaton-radion}.} where it was noted that this scenario is dual to a CFT with a marginal deformation by a composite operator $\lambda {\cal O}^\dagger {\cal O}$, where ${\cal O}$ is an $SU(2)_L$ doublet
operator.$^{[2]}$\footnotetext[2]{We thank Kaustubh Agashe for bringing this point to our attention.}
There are, however, key differences between our setup and the one presented in \cite{Vecchi}, which we will
discuss in section II.
In particular, we try to provide here a coherent account of this scenario from the AdS side.
We investigate the stabilization mechanism in this scenario
and discuss the level of fine-tuning involved.
We then find the couplings of the lowest scalar excitation of the
coupled SU(2) scalar doublet--gravity system.
More specifically, since this lowest excitation exists both in the metric and in the CP-even component of the bulk doublet,
this object has both the standard gravitational couplings of the metric excitation and the Yukawa and gauge couplings of the
bulk doublet.  We thus denote this mode the ``Higgs-radion", and upon calculating its couplings, we find that the ``Higgs-radion"
interpretation of the 126 GeV scalar is consistent with the LHC data for a KK gluon mass of ${\cal O}(5~TeV)$, which is not excluded by the latest experimental bounds.

\section{Radius Stabilization by an SU(2) bulk doublet}

In the scenario envisioned here there is a 5d SU$(2)_L$ bulk scalar doublet,
potentially with a VEV profile along the coordinate $y$ of the extra-dimension, within the general context of RSI \cite{RS}.
In this section we derive the solution to the Einstein equations of the coupled scalar-gravity system and find a set of
boundary conditions for which the EW-Planck hierarchy of scales can be generated without any large tuning apart from that which is required by the condition of a vanishing 4d cosmological constant. The main difference between our scenario and the GW mechanism for radius stabilization is that, both the size of the extra dimension and the mass of the EW gauge bosons depend on the VEV profile of our bulk doublet. In particular, as we will see below, the requirement of ${\cal O} (EW)$ masses will lead to a vanishingly small  VEV on the Planck brane,
as opposed to the case of the GW scalar \cite{GW,DeWolfe-GW,CSAKI-Radion}.

As usual, the metric is parameterized as:
\begin{equation}
ds^2=e^{-2A}dx^{\mu}dx^{\nu}\eta_{\mu \nu}-dy^2 ~,
\end{equation}
 where $A$ is the metric field which is determined by the Einstein equations
 (i.e., $A=ky$ for the $AdS_5$ solution, where $k$ is the curvature of the extra dimension).
 The two branes are located at $y=0$ (Planck Brane) and $y=y_c$ (TeV Brane).
Similar to the 4d Higgs mechanism, we define the 5d VEV as
\begin{equation}
\Phi=\left (
\begin{array}{c}
0 \\
\phi_0(y)
\end{array}
\right) ~.
\end{equation}

The profiles $A(y)$ and $\phi_0(y)$ are then determined by the Einstein equations.
We will assume that the backreaction, which is formally defined as $\ell$ later in the text (see Eq. \ref{backreaction}), is small and solve all
equations to the lowest non-vanishing order in $\ell$, unlike the full solution in \cite{DeWolfe-GW, CSAKI-Radion} and not using the effective ansatz of \cite{GW}.
We will later see that the assumption of a small backreaction is also required  to keep the EW scale lighter than the KK scale .

The bulk and brane actions are given by:
\begin{eqnarray}
S_{Bulk}&=&\frac{1}{2}\int{d^4x}\int_{0}^{y_c}{dy\sqrt{G}\left(G^{AB}\partial_A\Phi \partial_B \Phi-V(\Phi)+6\frac{k^2}{\kappa^2} \right)} ~, \\
S_{Brane}&=&-\int{d^4x}\sqrt{-g_i}V^{Brane}_i(\Phi) ~,
\end{eqnarray}
where
\begin{eqnarray}
V(\Phi)&=&m^2\Phi^2 ~, \label{bulkPotential}\\
V^{Brane}_i(\Phi)&=&\lambda_i\Phi^4+m_i^2\Phi^2+\Lambda_i ~,\\
\kappa^2&=&\frac{1}{2M^3_{Pl}} ~. \label{branePotential}
\end{eqnarray}
where as usual, $m,~m_i$ and $\lambda_i$ have mass dimensions 1, 0.5 and -2, respectively, while
$\Phi$ has a mass dimension 1.5.

The Einstein equations, $R_{ab}=\kappa^2 \tilde{T}_{ab}=\kappa^2\left(T_{ab}-\frac{1}{3}g_{ab}g^{cd}T_{cd} \right)$, give \cite{CSAKI-Radion}:
\begin{eqnarray}
4A'^2-A'' &=&4k^2-\frac{2\kappa^2}{3}V(\phi_0)-\frac{2\kappa^2}{3}V^{Brane}_i(\phi_0)\delta(y-y_i) ~, \\
A'^2 &=k^2+&\frac{\kappa^2 \phi^{\prime2}_0}{12}-\frac{\kappa^2}{6}V(\phi_0)  ~, \label{zeroenergy}
\end{eqnarray}
\begin{equation}
\phi^{\prime\prime}_0 = 4A'\phi^{\prime}_0+\frac{\partial V(\phi_0)}{\partial\phi_0}+\frac{\partial V^{Brane}_i(\phi_0)}{\partial\phi_0}\delta(y-y_i) ~, \label{field_eq}
\end{equation}
where primes denote $\partial_y$ and $V(\phi_0)=V(\Phi)$ for $\Phi=\left( \begin{array}{c c} 0 \\ \phi_0 \end{array}\right)$. The boundary conditions are
then given by matching the delta functions:
\begin{eqnarray}
\left[ \phi_0^\prime  \right]_i=\frac{\partial V^{Brane}_i(\phi_0)}{\partial\phi_0} ~, \label{const1}\\
\left[ A' \right]_i=\frac{\kappa^2}{3}V^{Brane}_i(\phi_0) \label{const2}~.
\end{eqnarray}

It is now vital to understand what is the number of integration constants and constraints.
There are two integration constants for $\phi_0$, as can be seen by inserting Eq. \ref{zeroenergy}
into Eq. \ref{field_eq}, so that the resulting equation is a second order differential equation.
The first derivative of $A$ is then completely determined from Eq. \ref{zeroenergy},  with no new integration constants
 (the value of $A$ does not enter the equations and so it is irrelevant).
Thus, after setting the integration constants for $\phi_0$ as discussed above, we are left with two additional boundary conditions (i.e., out of the four boundary conditions in Eqs.~\ref{const1} and \ref{const2}).
One of them will determine the radius of the extra dimension
  and the remaining one will have to be fine-tuned.
  This fine-tuning corresponds to the cosmological constant problem, which in the RS framework
  appears as the fine-tuning of the Planck brane tension.
  Recall that without the bulk scalar, there are only two boundary conditions and no relevant integration constants
  so that both of them have to be fine-tuned.

Using Eq.~\ref{zeroenergy} and the boundary conditions for $\phi_0^\prime$ in Eq.~\ref{const1},
the two boundary conditions for $A^\prime_i$ in Eqs.~\ref{const2}
can be rewritten as follows:
\begin{equation}
\left(\frac{\kappa^2}{6}V^{Brane}_i(\phi_0)\right)^2 = k^2+\frac{\kappa^2}{24}
\left(\frac{\partial V^{Brane}_i(\phi_0)}{\partial\phi_0}\right)^2-\frac{\kappa^2}{6}V(\phi_0) \label{phi_boundary} ~,
\end{equation}
which, as Eq.~\ref{const1}, has to be satisfied on each of the two branes. Notice that
no derivatives of the VEV profile (i.e., no $\phi_0^\prime$) appear in Eq.~\ref{phi_boundary},
so that it is simply an equation for the value of $\phi_0$ on the two branes ($i=TeV,~Pl$),
of the form $f(\phi^i_0)=0$.
The natural values for $\phi_0^i$ will thus be
of the order of the Planck scale, which is phenomenologically unacceptable, since the effective 4d VEV which sets the
EW-scale (gauge-bosons masses) is sensitive to its profile (see Eq.~\ref{veff0}). We will return to
this apparent problem below.

Due to the fact that Eq.~\ref{phi_boundary} just sets the values of $\phi_0$ on the two branes,
we redefine it to be of the form:
\begin{equation}
\phi_0|_i \equiv \phi_{TeV/Pl} \label{phi_boundary_2}~,
\end{equation}
where $\phi_{TeV}$ and $\phi_{Pl}$ are the solutions  to Eq.~\ref{phi_boundary}
for $\phi_0$ on the TeV and Planck branes, respectively.

\begin{figure}[htb]
\begin{center}
\includegraphics[scale=0.6]{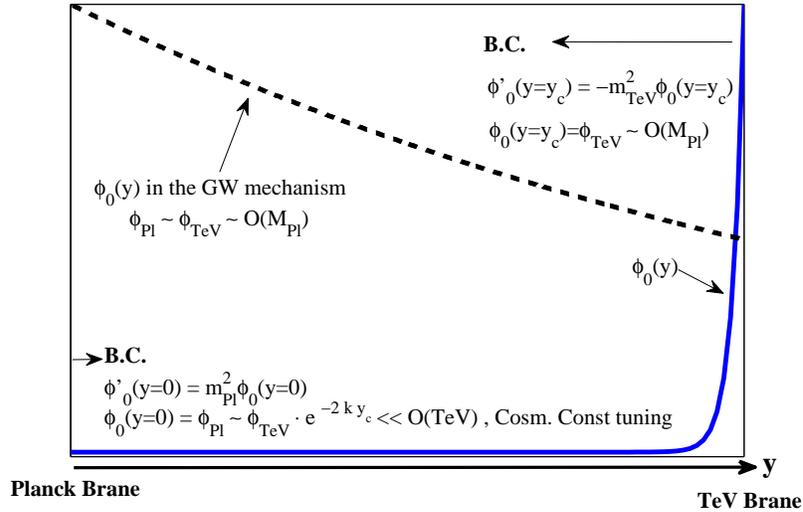}
\end{center}
\caption{\emph{A graphic illustration of our boundary conditions on $\phi_0$ and $\phi_0^\prime$
on the TeV and Planck branes (solid line). A typical solution of the VEV profile in the GW mechanism 
is also shown (dashed line) for comparison
(see also Table \ref{tab0} and discussion in the text).}
\label{fig1}}
\end{figure}

Assuming that the backreaction $\ell$ is small so that $A'=k+{\cal O}(\ell^2)$ (see Eq.~\ref{backreaction}),
Eq.~\ref{field_eq} in the bulk becomes:
\begin{equation}
\phi^{\prime\prime}_0 = 4k\phi^{\prime}_0+\frac{\partial V(\phi_0)}{\partial\phi_0} ~,
\end{equation}
with the general solution
\begin{equation}
\phi_0=e^{2k(y-y_c)}\left(C_1e^{\nu k(y-y_c)}+C_2e^{-\nu k(y-y_c)}\right) \label{phi_solution} ~,
\end{equation}
where $\nu=\sqrt{4+m^2/k^2}$ and $m$ is the mass of the bulk doublet, as defined in Eq.~\ref{bulkPotential}.
This is the point where our analysis deviates from the ``conventional" GW mechanism. In particular,
if the wave-function of the gauge-boson zero modes is flat, which will turn out to be a good approximation in our
case (see Eq.~\ref{veff} and discussion thereof), then the effective EW VEV is given by:
\begin{equation}
v^2_{eff}=\int\limits_y \ \phi^2_0e^{-2ky} \label{veff0} ~,
\end{equation}
which in general gives $v_{eff}\sim {\cal O}(M_{Pl})$, unless
$\phi_0(y=0)/M_{Pl} \ll 1$.
For our solution, assuming no tuning in the values of $C_1$ and $C_2$ (so that they are both of ${\cal O}(M_{Pl})$),
the condition of a vanishing $\phi_0$ on the Planck brane (i.e., in the sense that $\phi_0(y=0)/M_{Pl} \to 0$) is obtained when $\nu<1$ and $\phi_0=C_2e^{(2-\nu)k(y-y_c)}$ for
$y<<y_c$. This is different from the GW mechanism
in which $\nu \sim 2$ \cite{GW}, resulting in an
${\cal O}(M_{Pl})$ VEV on the Planck brane (i.e., $\phi_0(y=0)|_{GW} \sim {\cal O}(M_{Pl})$).
It is, therefore, clear that the GW mechanism, as it is, cannot be applied with an $SU(2)_L$
doublet stabilizer.
Since the GW mechanism cannot work with $\nu < 1$,
we adopt a different choice for
applying the boundary conditions and for the fine-tuning
condition required in order to have a vanishing cosmological constant.
In particular, in the GW mechanism
the values of $\phi_0$ are imposed on both branes (as in Eq.~\ref{phi_boundary_2}),
while one of the conditions on the
derivative of $\phi_0$ (Eq.~\ref{const1}) is left to be tuned (see e.g., \cite{DeWolfe-GW}).
This is clearly not a choice we can make, since in choosing $\phi_0(y=0)=\phi_{Pl}$, we expect the
natural value on the Planck brane to be
$\phi_{Pl}\sim {\cal O}(M_{Pl})$, which is exactly what we are trying to avoid.
We, therefore, use instead the two boundary conditions for the derivatives $\phi_0^\prime$ on both branes (Eq.~\ref{const1})
and the boundary condition for $\phi_0$ on the TeV brane
(i.e., $\phi_{TeV}$ in Eq.~\ref{phi_boundary_2}). As we will shortly see, the remaining fourth condition for $\phi_{Pl}$ in Eq.~\ref{phi_boundary_2} will have the usual fine-tuning associated with the cosmological constant.
Nonetheless, the condition $\phi_{Pl}/M_{Pl} \to 0$ which follows from the choice $\nu < 1$ (see above) is not in
conflict with our choice of boundary conditions and, as is shown below, it does not add to the level of fine-tuning required for
the cosmological constant problem. A graphic illustration of the set of boundary conditions in our setup is given in
Fig.~\ref{fig1}.

Inserting now the solution for $\phi_0$ (Eq.~\ref{phi_solution}) into the boundary conditions of Eqs.~\ref{const1}
and \ref{phi_boundary_2}, and ignoring for now the condition $\phi_0(y=0)=\phi_{Pl}$, we obtain:

\begin{eqnarray}
\left(C_1+C_2\right) &=& \phi_{TeV}\label{phi0} ~,
\end{eqnarray}
\begin{eqnarray}
k(C_1(\nu+2)e^{k(2+\nu)(y_i-y_c)}&+&C_2(2-\nu)e^{k(2-\nu)(y_i-y_c)})
=\pm \left(2\lambda_i\phi(y_i)^3+m_i^2\phi(y_i) \right)\label{der_cond} ~,
\end{eqnarray}
where the $\pm$ in Eq.~\ref{der_cond} corresponds to the Planck/TeV brane, respectively.

In particular, using $\phi_0(y=0) << m_{Pl}$ on the Planck brane
(see discussion above),
we can neglect the cubic term $2\lambda_{Pl} \phi(y=0)^3$ in Eq.~\ref{der_cond}, obtaining:
\begin{eqnarray}
C_2 &=& \eta_1 C_1 e^{-2k\nu y_c} ~, \\
\eta_1 &\equiv& \frac{(2+\nu-m_{Pl}^2/k)}{(-2+\nu+m_{Pl}^2/k)} ~,
\end{eqnarray}
and using Eq.~\ref{phi0}, also
\begin{eqnarray}
C_1=\phi_{TeV}\frac{1}{1+\eta_1 e^{-2\nu k y_c}} ~.
\end{eqnarray}

The third boundary condition sets the radius of the extra dimension $y_c$:
\begin{eqnarray}
y_c &=& \frac{1}{2k\nu}log \left(\frac{\eta_1}{\eta_2}  \right) \label{stabrad} ~,
\end{eqnarray}
with
\begin{eqnarray}
\eta_2 &\equiv& \frac{(2+\nu+2\lambda_{TeV}\phi_{TeV}^2+m^2_{TeV}/k)}{(-2+\nu-
2\lambda_{TeV}\phi_{TeV}^2-m^2_{TeV}/k)}\approx \frac{(2+\nu+m^2_{TeV}/k)}{(-2+\nu-m^2_{TeV}/k)} \label{eta2}~,
\end{eqnarray}
where the last approximation is applied due to a small backreaction (i.e., $\kappa\phi_{TeV}<<1$, see
Eq.~\ref{Aprime}) in which case we can
neglect the contribution from the quadratic terms $\propto \lambda_{TeV}\phi_{TeV}^2$.

Recall that the hierarchy of scales in RS setups is given by $\frac{IR-scale}{Pl-scale} \sim e^{-k y_c}$,
which, using Eq.~\ref{stabrad}, gives in our case
$\frac{IR-scale}{Pl-scale} \sim e^{-\frac{1}{2\nu}} =
e^{-\left({2 \sqrt{4+\frac{m^2}{k^2}}} \right)^{-1}}$.
This is in contrast to the GW case where
$\frac{IR-scale}{Pl-scale} \sim
e^{-4 \frac{k^2}{m^2}}$.
Thus, the required hierarchy of scales is obtained in our case when
$\frac{1}{2\nu} \sim 30$ (i.e., when $\frac{m^2}{k^2} \to - 4 + {\cal O}(0.001)$), while in the GW case the condition is
$4 \frac{k^2}{m^2} \sim 30$ (i.e., $\frac{m^2}{k^2} \to  {\cal O}(0.1)$).

The remaining fourth boundary condition, $\phi(y=0)=\phi_{Pl}$ in Eq.~\ref{phi_boundary_2}, 
which we have omitted from the above discussion, naively seems to involve
a severe fine-tuning to ensure $\phi_{Pl} < {\cal O}(TeV)$.
In order to understand this apparent problem,
let us re-write the original equation of the boundary condition for $\phi_0(y=0)$ explicitly (see Eq.~\ref{phi_boundary}):
\begin{equation}
\left(\frac{\kappa^2}{6}V^{Brane}_{Pl}\left(\phi_0(y=0)\right)\right)^2 = k^2+\frac{\kappa^2}{24}\left(\frac{\partial V^{Brane}_{Pl}\left( \phi_0(y=0) \right)}{\partial\phi_0}\right)^2-
\frac{\kappa^2}{6}V \left( \phi_0(y=0) \right)  \label{Plancktune} ~,
\end{equation}
which is generically of the form
\begin{equation}
\sum\limits_{i=2}^8a_i\phi_{Pl}^i-\Lambda^2_{Pl}\frac{\kappa^4}{36}+k^2=0
\label{4BC}~.
\end{equation}

As we have shown above, in order to have a phenomenologically acceptable framework, $\phi_0(y=0) = \phi_{Pl}$
has to be significantly smaller than ${\cal O}(TeV)$, 
or more specifically $\phi_0(y=0) \approx \phi_{TeV} e^{-2ky_c}$ (see Eq. \ref{phi_solution}).
 This seems to require fine-tuning between the ${\cal O}(M_{Pl}^2)$ terms
 $\propto \Lambda_{Pl}^2,k^2$ in Eq.~\ref{4BC}.
 In particular, these constant ${\cal O}(M_{Pl}^2)$ terms have to cancel up to  $\sum\limits_{i=2}^8a_i\phi_{Pl}^i < {\cal O}(TeV^2)$.
 However since there are no additional free parameters left,
 this equation has to be exactly tuned in any case (i.e.,
 even when $\sum\limits_{i=2}^8a_i\phi_{Pl}^i \sim {\cal O}(M_{Pl}^2)$)
 in order to ensure a vanishing 4d cosmological constant - this is
 exactly the cosmological constant problem plaguing any RS setup.
 The level of fine-tuning required in the general case
 is, therefore, $\sim 122$ orders of magnitude, far greater than the fine-tuning required by any ${\cal O}(TeV)$ terms in this equation.
 Thus, the condition/requirement (in our case) that
 $\phi_{Pl} < {\cal O}(TeV)$ has no effect on the level of fine-tuning required for the cosmological constant problem.
 In turn, in our scenario, this seems to imply that whatever solves the cosmological constant problem should be
 manifest on the Planck brane or in its vicinity - a reasonable assumption from the phenomenological perspective.
 We thus find that the only fine-tuning required in our scenario is that of the cosmological constant which is inherent in any
 RS construction with a stabilized radius (as in the GW mechanism).

In order to have a solution for $y_c$ we need $\eta_1>\eta_2>0$ (see Eq.~\ref{stabrad}). It was already stated that in order to generate the desired EW-Planck hierarchy,
$\nu$ has to be small but not hierarchically small. It is easy to verify that for $2-\nu<-m^2_{TeV}/k<m^2_{Pl}/k< 2+\nu$ the condition of  $\eta_1>\eta_2>0$ is satisfied, with only a small amount of tuning involved.
We further make the simplifying assumption that $\eta_1 \gg 1$, which is satisfied if $m^2_{Pl}/k\approx 2-\nu$, so that
\begin{equation}
\phi_0(y)=\phi_{TeV}e^{(2-\nu)k(y-y_c)} \label{phi0def} ~.
\end{equation}

As mentioned above, we take the backreaction  to be small, or equivalently take $\kappa \phi_{TeV}\ll 1$ (see the second term in Eq.~\ref{Aprime} below).
Under this assumption, the metric field $A$ can be calculated from Eq.~\ref{zeroenergy} :
\begin{eqnarray}
A' &=& k+\frac{\kappa^2 \phi^{\prime2}_0}{24k}-\frac{\kappa^2}{12k}V(\phi_0)= \nonumber \\
&=&k+ \frac{1}{24} e^{(2-\nu)2k(y-y_c)} \kappa^2 \phi_{TeV}^2 (20 - 4 \nu - 3 \nu^2)k \equiv k\left(1+\frac{1}{6} e^{-2uy} \ell^2     \right) \label{Aprime}~,
\end{eqnarray}
where
\begin{eqnarray}
u &\equiv& (\nu-2)k ~,\\
\ell^2 &\equiv& \frac{1}{4} e^{-(2-\nu)2ky_c} \kappa^2 \phi_{TeV}^2 (20 - 4 \nu - 3 \nu^2) ~, \label{backreaction}
\end{eqnarray}
so that
\begin{eqnarray}
\phi_0(y) &=& \phi_{TeV}e^{-u(y-y_c)} \label{phi0eq} ~,
\end{eqnarray}
and $\ell$ is now parameterizing the backreaction. In particular,
$A'=k+ {\cal O}(\ell^2)$ at lowest order,
consistent with our assumption above.

Let us give a brief account of the free parameters involved in our setup, i.e., the parameters $k,~m,~\lambda_{TeV},~\lambda_{Pl},~m^2_{TeV}$, $m^2_{Pl},~\Lambda_{TeV},~\Lambda_{Pl}$, which appear in
the scalar potentials in the bulk and on the two branes.
As we have seen, $\lambda_{TeV}$ and $\lambda_{Pl}$ do not enter the solution in the small backreaction approximation (see Eq.~\ref{der_cond}). Also, the condition of a stabilized radius requires $2-\nu<-m^2_{TeV}/k<m^2_{Pl}/k< 2+\nu$, which leads to $m^2_{Pl}/k \approx -m^2_{TeV}/k \approx 2$.
The size of the extra dimension, $y_c$, which is given
as a function of $k,~m,~\lambda_{TeV},~\lambda_{Pl},~m^2_{TeV}$
and $m^2_{Pl}$ (Eq.~\ref{stabrad}), sets the bulk mass parameter $m$
by requiring that it reproduces the desired hierarchy between the EW and Planck scales,
i.e., $k y_c \sim 30$, leading to
$m^2 \to - 4 k^2$ (see discussion below Eq.~\ref{eta2}).
As for the brane tensions, $\Lambda_{Pl}$ is set by the cosmological constant tuning condition (Eq.~\ref{Plancktune}), while $\Lambda_{TeV}$ is traded with the value of
$\phi_0$ on the TeV brane, i.e.,  $\phi_0(y=y_c) \equiv \phi_{TeV}$.
We are thus left with three parameters: $k, y_c$ and  $\phi_{TeV}$ and the KK-scale will be
determined by the exact value of $k y_c$. In particular,
$k y_c \sim 30$ will be required in order to keep the KK-scale
at the TeV range.

It is also instructive
to consider the dual CFT picture that corresponds to our scenario.
As noted in the introduction, the doublet scalar on the AdS side is dual to a doublet operator ${\cal O}$ on the CFT side, giving \cite{Vecchi}:
\begin{equation}
L_{\cal O}=\lambda {\cal O}^\dagger {\cal O}~,
\end{equation}
which is  marginal when $dim({\cal O}) \approx 2$.
In particular, under the AdS/CFT duality, $dim({\cal O})=2+\nu$,
so that  the condition we have just derived for the EW-Planck hierarchy, i.e, that $\nu<<1$, is the same as the condition that the dual operator in the CFT picture is marginal. For comparison, in the
GW mechanism the singlet scalar is dual to a singlet operator appearing linearly in the lagrangian $L_{\cal O}=J{\cal O}$. The marginal dimension for such an operator is $dim({\cal O})=4$, which is satisfied on the AdS side when $\nu=2$. Thus, the condition in the GW mechanism that $m^2/k^2 << 1$, is responsible for both
the generation of the EW-Planck hierarchy and the marginality of the CFT deformation.

\section{The Higgs-Radion Mass}

Let us now consider the fluctuations about the above (classical) background. We will closely follow the discussion in \cite{CSAKI-Radion}. It is easy to verify that the goldstone modes  are not relevant, as only the first derivative of the scalar potentials enters the Einstein equations \cite{CSAKI-Radion}. The relevant scalar and metric excitations, denoted here as $\varphi(y,x)$ and $F(y,x)$, are thus defined as follows:
\begin{eqnarray}
\Phi&=&\left (
\begin{array}{c}
0 \\
\phi_0(y)+\varphi(y,x)
\end{array}
\right) \label{phidef} ~, \\
ds^2&=&e^{-2A-2F(y,x)}dx^{\mu}dx^{\nu}\eta_{\mu \nu}-(1+2F(y,x))^2dy^2 ~.
\end{eqnarray}

One can then derive, from the Einstein equations, the coupled wave equations for both excitations
 \cite{CSAKI-Radion}:
\begin{eqnarray}
F''&-&2A'F'-4A''F-2\frac{\phi''_0}{\phi'_0}F'+4A'\frac{\phi''_0}{\phi'_0}=e^{2A}\Box F \label{WaveF} ~, \\
\phi_0'\varphi&=&\frac{3}{\kappa^2}\left(F'-2A'F \right) \label{Wavephi} ~,
\end{eqnarray}
which implies
that the physical fluctuations $\varphi$ and $F$ correspond in fact to the same state.

The KK expansion of the coupled system can be written
as
\begin{eqnarray}
\varphi(x,y)&=&\sum \varphi_n(y)h_n(x) \label{KKphi} ~,\\
F(x,y)&=&\sum F_n(y)h_n(x) \label{KKF} ~,
\end{eqnarray}
where each KK mode in the above KK expansion satisfies $\Box h_n=-m_n^2h_n$, so that Eq.~\ref{WaveF}, with the solution for $\phi_0$  in Eq.~\ref{phi0eq}, gives
\begin{equation}
F_n''-2A'F_n'-4A''F_n+2uF_n'-4uA'F_n+m_n^2e^{2A}F_n=0 \label{WaveFfull} ~.
\end{equation}

Then, to the lowest non vanishing order of the backreaction, the solution for the zero mode, which is our  ``Higgs-radion" state, is given by \cite{CSAKI-Radion}:
\begin{eqnarray}
F_0 &=& e^{2ky}(1+\ell^2f_0(y)) \label{F0} ~, \\
m^2_{h_r} &\equiv& m_0^2 = \ell^2 \tilde{m}_0^2 \label{mrad} ~, \\
f'_0(y) &=& Ce^{-2(k+u)y}-\frac{\tilde{m}_0^2}{2(2k+u)}e^{2ky}
-\frac{2(k-u)u}{3k}e^{-2uy} \label{fsolution} ~,
\end{eqnarray}
where $m_{h_r} = m_0$ is the mass of the lowest excitation in the KK tower of $\varphi$ and $F$ (Eqs.~\ref{KKphi} and \ref{KKF}), i.e.,
of our Higgs-radion, and together with the integration constant $C$, it is determined by the boundary conditions
(see below). Note that in the solution to Eq.~\ref{WaveFfull},  $m_{h_r}$ is linear with the backreaction $\ell$.
We therefore define it using $\tilde{m}_0$,
which is independent of the backreaction.

The relevant boundary conditions in our case are a bit different from the ones used in \cite{CSAKI-Radion}, as we cannot make the simplifying assumption $\lambda_{TeV/Pl} \gg 1$. In particular,
they are:
\begin{equation}
\left[\varphi' \right]_i=\frac{\partial^2 V^{Brane}_i(\phi_0)}{\partial \phi^2} \varphi + 2 \frac{\partial V^{Brane}_i(\phi_0) }{\partial \phi}F
\label{BCF1}~.
\end{equation}

Recall that the condition of a stabilized radius
requires $2-\nu<-m^2_{TeV}/k<m^2_{Pl}/k < 2+\nu$, which for
a phenomenologically viable size of the radius (i.e., when
$\nu \ll 1$)
leads to  $m^2_{Pl}/k \approx -m^2_{TeV}/k \approx 2$.
Thus the solution to Eq.~\ref{BCF1}, to lowest order in
$\ell^2$ and $\nu$, is given by:
\begin{equation}
m^2_{h_r} \approx \ell^2 \frac{52k^2}{15 ky_c} e^{2ky_c} \label{mr}~,
\end{equation}
and, under the same assumptions, the effective 4d VEV is ($k y_c \sim 30$):
\begin{equation}
v_{eff}^2 \approx \ell^2 \frac{2}{5k \kappa^2}e^{2ky_c} \label{veff2} ~.
\end{equation}
We thus find that the ratio between the Higgs-radion mass and
the 4d VEV is:
\begin{equation}
\frac{m^2_{h_r}}{v_{eff}^2} \approx \frac{26}{3 ky_c} k^3 \kappa^2 \approx \frac{1}{4} k^3 \kappa^2 ~,
\end{equation}
so that, for $v_{eff}=246/\sqrt{2}$ GeV and $m_{h_r}=126$ GeV, the curvature of the extra dimension is determined (up to some theoretical uncertainty): $k/M_{Pl} \approx 1.6$, which is within the validity range of this ratio, see \cite{Warped-Grav}. In addition, Eqs.~\ref{mr}
and \ref{veff2} can be used to express $\ell^2$ as a function of
$y_c$ as well, from which $\phi_{TeV}$ can be extracted (see
Eq.~\ref{backreaction}). We are, therefore, left
with $y_c$ as the only free parameter which will determine the KK
scale (see next sections).
\begin{figure}[htb]
\begin{center}
\includegraphics[scale=0.4]{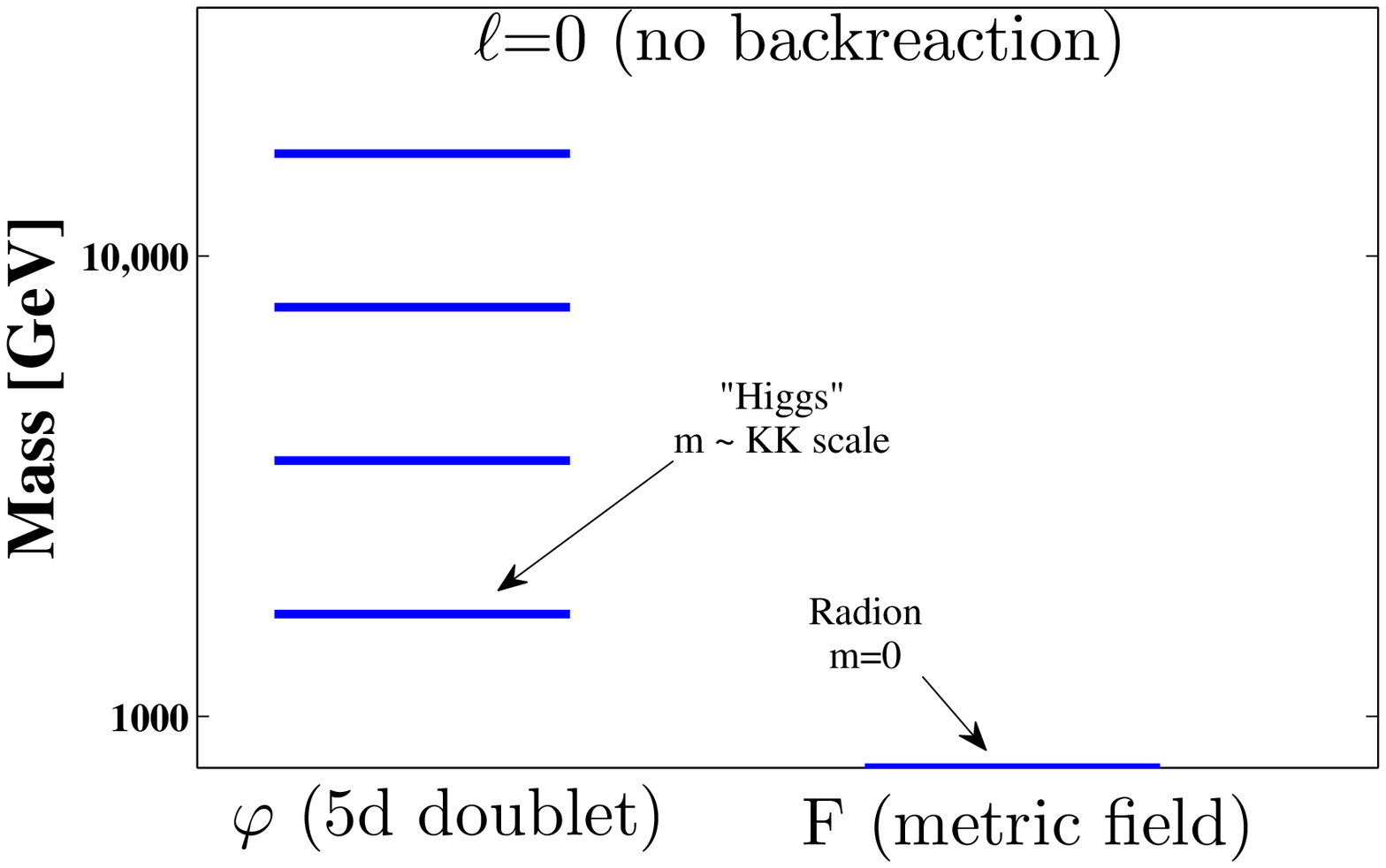}
\includegraphics[scale=0.4]{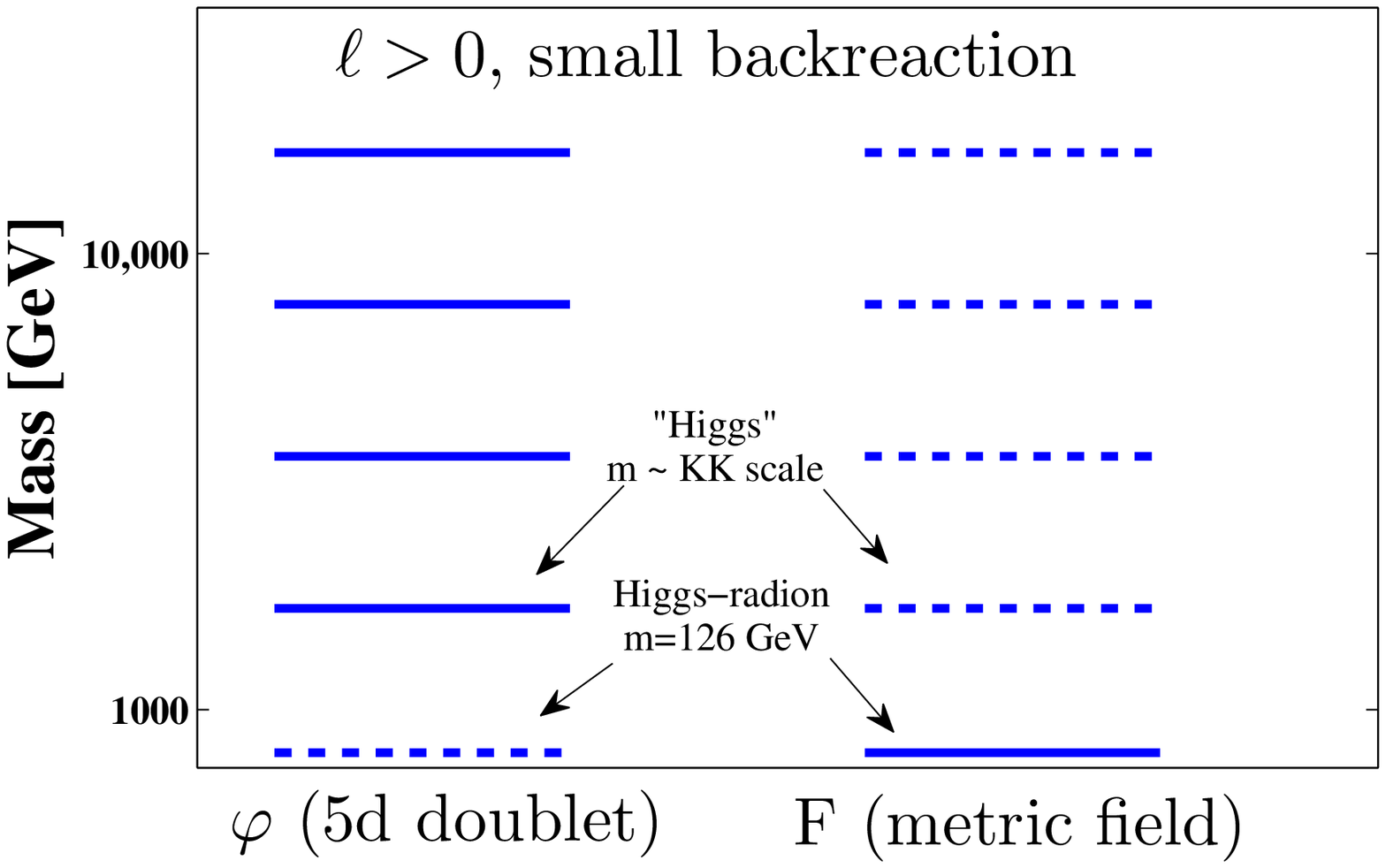}
\end{center}
\caption{\emph{A graphic illustration of the particle/KK spectrum in our setup with (right) and
without (left) backreaction.}
\label{fig2}}
\end{figure}

Let us briefly address the higher excitations of this system. Recall that when the gravitational effect is turned off, i.e., the backreaction is taken to zero, we are left with only the KK tower in $\varphi$, which
 is independent of $F$ for which only the zero mode remains \cite{CSAKI-Radion}. The case of no backreaction may correspond to a pure RS scenario (i.e., no stabilization) or to a stabilization by some other mechanism, where the condition in Eq.~\ref{Wavephi} is saturated by some other scalar (e.g., the GW mechanism).
Thus, in this case, the massless zero mode of $F$ is the radion
and the lowest excitation in $\varphi$ is identified with the ``Higgs".
Once the backreaction is reintroduced, this ``Higgs" state
becomes the second KK excitation of the $F-\varphi$ system
(denoted below as $\varphi_1$) and
our Higgs-radion state is then the lowest KK excitation
of the coupled/backreacted $F-\varphi$ system.
It should be stressed that this Higgs-radion state is NOT the conventional
mixed state between the GW radion and the ``Higgs", which was studied in \cite{Higgs-Radion-Mixing,Higgs-Radion-Mixing-2}.
{\small
\begin{table}[htb]
\begin{center}
\begin{tabular}{|c||c|c|}
\hline
 & GW mechanism & Our setup \\
 \hline \hline
Stabilizing field & scalar singlet & SU(2) scalar doublet \\
\hline
The bulk mass parameter $\left( V(\Phi) = m^2 \Phi^2 \right) $ & $m^2 \ll 1$ & $m^2 \to -4k^2$ \\
\hline
VEV profile, $\phi_0(y)$& nearly flat & steep,  \\
 & & peaked on the TeV brane \\
 \hline
 TeV brane VEV, $\phi_{TeV} \equiv \phi_0(y=y_c)$ & $\phi_{TeV} \sim {\cal O}(M_{Pl})$ & $\phi_{TeV} \sim {\cal O}(M_{Pl}) $ \\
\hline
Planck brane VEV, $\phi_{Pl} \equiv \phi_0(y=0)$ & $\phi_{Pl} \sim {\cal O}(M_{Pl})$ & $\phi_{Pl} \sim M_{Pl}e^{-2ky_c} \ll {\cal O}(eV)$  \\
\hline
Lowest scalar excitation & Radion & Higgs-radion \\
\hline
& & both metric couplings  \\
(Higgs-)Radion couplings & purely metric couplings  & and Yukawa/Gauge couplings  \\
& & of the doublet  \\
\hline
\end{tabular}
\caption{A summary of the most notable differences between our setup and the GW mechanism.}
\label{tab0}
\end{center}
\end{table}
}

Unlike the mass of our Higgs-radion state, the mass of what we have denoted as the ``Higgs" state ($\varphi_1$) does not vanish for $\ell \to 0$, so that the backreaction can be neglected
when evaluating its mass. Numerically we find that
$m_{\varphi_1} \approx 3.38 k e^{-ky_c}$, which is of the order of the
KK-scale. It is therefore evident that, in our model, 
the only light state in the scalar spectrum is the Higgs-radion,
while all other states have masses well above 1 TeV.
A graphic illustration of the particle/KK spectrum in our setup with and
without backreaction is given in Fig.~\ref{fig2}.

In Table \ref{tab0} we
summarize the notable differences between our setup and the GW mechanism for radius stabilization.

\section{Higgs-Radion Interactions}
Let us explore now the interactions of our Higgs-radion with the SM fields. 
This requires the Higgs-radion field to be canonically normalized.
In particular, we have seen that (see Eqs.~\ref{Wavephi} and \ref{F0}):
\begin{eqnarray}
F_0&=&e^{2ky}\\
\varphi_0&=&\frac{3}{\kappa^2\phi_0'}\left( F_0'-2A'F_0 \right) ~,
\end{eqnarray}
where we have taken only the leading order term in the backreaction for $F_0$.
Defining $h_0(x)$ as:
\begin{equation}
h_0(x)=\frac{h_r}{N} ~,
\end{equation}
 so that $h_r$ is the canonically normalized Higgs-radion field,
the normalization factor $N$ can be calculated from:
\begin{eqnarray}
N^2&=&N^2_F+N^2_{\varphi} ~,
\end{eqnarray}
where
\begin{eqnarray}
N^2_{\varphi} &=& \int e^{-2ky}\varphi^2 dy \approx 6 \ell^2M_{Pl}^2e^{6ky_c}  \approx 30 v_{eff}^2 e^{4ky_c} ~,\\
N^2_{F} &=&  6M_{Pl}^2e^{2ky_c} ~.
\end{eqnarray}

In particular, notice that the contribution to the kinetic term coming from the metric field $F$ is the same as in \cite{CSAKI-Radion,CSAKI-REAL}. We also see that the contribution to the kinetic term coming from $\varphi_0$ is ${\cal O}(\ell^2)$ as in \cite{CSAKI-Radion} (but without an additional parametrical suppression that is present in \cite{CSAKI-Radion}).
The normalization is then:
\begin{eqnarray}
N^2&\approx&N_F^2=(\Lambda_r e^{2ky_c})^2 ~, \\
\Lambda_r &\equiv& \sqrt{6}M_{Pl}e^{-ky_c}~,
\end{eqnarray}

and the 5d fields $F(x,y)$ and $\varphi(x,y)$ are therefore given by:
\begin{eqnarray}
F(x,y) &=& h_r\frac{e^{2k(y-y_c)}}{\Lambda_r} +KK(n>0) ~, \\
\varphi(x,y) & \approx & h_r\frac{1}{\Lambda_r e^{2ky_c}} \left( 2\sqrt{5} e^{4ky} \sqrt{\frac{1}{\kappa^2}}-\frac{13 e^{2 k y + 2 k y_c} \sqrt{\frac{1}{\kappa^2}} y}{\sqrt{5} y_c} \right) \ell +KK(n>0) ~.
\end{eqnarray}

We see that both $F(x,y)$ and $\varphi(x,y)$ have an $h_r$
(the Higgs-radion) component, so that the couplings of our
Higgs-radion are induced by both the gravitational couplings of $F(x,y)$ and the couplings of $\varphi(x,y)$ - specifically its gauge and Yukawa couplings. Note that the gravitational couplings of $F(x,y)$ are similar to the ones of the pure radion in the conventional RS models with the GW mechanism \cite{CSAKI-Radion,CSAKI-REAL}.
Also, as we will shortly see, the couplings of the Higgs-radion to the gauge-bosons and to
the fermions are not negligible despite the fact that the $h_r$ component in
$\varphi(x,y)$ is ${\cal O}(\ell)$.
Thus, in contrast with \cite{Vecchi}, the phenomenology of our Higgs-radion is different from the pure radion case.

 Let us consider the couplings of $h_r$ to the top-quark and the gauge-bosons, as these are the ones relevant for the LHC phenomenology of the our Higgs-radion.
Naturally, we set the mass of the Higgs-radion, $h_r$, to be 126 GeV. For the coupling to the top, we assume that the right-handed top, $t_R$, is localized on the TeV brane (this is not the case for the light fermions). Thus, the coupling of the Higgs-radion to the top-quark, coming from the couplings of the metric field $F$ to TeV localized matter, is \cite{CSAKI-REAL}:
\begin{equation}
 L^F_{h_rtt}=\frac{h_r}{\Lambda_r}T^{\mu}_{\mu} \left|_{t \bar t} \right.=\frac{h_r}{\Lambda_r} m_t \overline{t}t
\end{equation}
while another contribution to its coupling to the top-quark arises from the 5d Yukawa couplings of $\varphi$ \cite{RS-flavor}:
\begin{eqnarray}
L^\varphi_{h_rtt}&=-&\int dy  \sqrt{G}y^{5d}_{t}\Phi \bar{t_L} t_R \label{varphitop}~,
\end{eqnarray}
where
\begin{eqnarray}
t_{R}=\frac{1}{\sqrt{N_{t1}}}\delta(y-y_c) ~,~
t_{L}(y=y_c)= \frac{1}{\sqrt{N_{t2}}} ~.
\end{eqnarray}

Also, $y^{5d}_t$ is the 5d Yukawa coupling of the top, $t_R$ and $t_L$ are the wave functions of the left and right handed top, and $N_{ti}$ are the corresponding normalization factors. The mass of the top is
 proportional to $<\Phi> = \phi_0(y=y_c)$ and is given by:
\begin{equation}
m_t=\frac{y^{5d}_t}{N_{t1}N_{t2}}\phi_{TeV}e^{-4ky_c} ~.
\end{equation}

The coupling of the $\varphi$ component of the Higgs-radion to the top-quark is obtained from Eq.~\ref{varphitop}:
\begin{eqnarray}
 L^\varphi_{h_rtt} &=& -\frac{y^{5d}_{t}e^{-4ky_c}}{N_{t1}N_{t2}}h_r\overline{t} t \ \int \varphi_0(y) \delta(y-y_c) dy  \nonumber \\
 &=& -\frac{y^{5d}_{t}e^{-4ky_c}}{N_{t1}N_{t2}}h_r\overline{t} t \ \frac{1}{\Lambda_r e^{2ky_c}} \left(2 \sqrt{5} e^{4ky_c} \sqrt{\frac{1}{\kappa^2}}-\frac{13 e^{4 k y_c} \sqrt{\frac{1}{\kappa^2}} }{\sqrt{5}} \right) \ell \nonumber \\
 &=&3\frac{h_r}{\Lambda_r} m_t \overline{t}t ~,
\end{eqnarray}
so that the overall Higgs-radion coupling to the top-quark is
\begin{equation}
L_{h_rtt}=\frac{4}{\Lambda_r} m_t h_r\overline{t}t ~.
\end{equation}

Next we calculate the couplings of our Higgs-radion to the massive gauge-bosons.
In order to do that, we first solve the equation of motion
for the gauge-boson zero modes $V_0$ \cite{Gauge-ADS5}:
\begin{equation}
\partial_y \left( e^{-2ky_c} \partial_y V_0 \right)-\frac{g^2_5}{4} \phi_0^2(y) e^{-2ky_c}V_0+m_n^2V_0=0 ~,
\end{equation}
where, $g_5$ is the 5d SU(2) gauge coupling, related to the 4d gauge coupling via $g_4=\frac{g_5}{\sqrt{y_c}}$ (see \cite{Gauge-ADS5}).
We treat the second term in the above equation (proportional to
$\phi_0^2 \sim \ell^2)$ as a perturbation and obtain $V_0=\frac{1}{\sqrt{y_c}}(const+V^{pert}_0)$.
Under the boundary conditions that the derivatives vanish on the branes (as there are no delta functions in the equation), we find
that
\begin{equation}
m_W=\frac{g_5}{2\sqrt{y_c}}v_{eff}=\frac{g_4}{2}v_{eff} \label{veff}~,
\end{equation}
so that $v_{eff}$ is indeed the equivalent of the EW VEV of the SM. That
is, when $v_{eff}=246/\sqrt{2}$ GeV we reproduce the correct mass for the W boson. The same holds for the $Z$-boson.

The dominant coupling of the Higgs-radion to the massive gauge-bosons from the metric field $F$ is \cite{CSAKI-REAL} (we drop the subscript
0 in $V_0$):
\begin{equation}
L^F_{h_rVV}=-\frac{h_r}{\Lambda_r} m^2_VV^\mu V_\mu ~,
\end{equation}
where $V^\mu V_\mu = 2 W^+_\mu W^{\mu-}$ for the W-boson and $V^\mu V_\mu=Z_\mu Z^\mu$ for the Z-boson.
The coupling coming from $\varphi$ (from the 5d covariant derivative of V)
is:
\begin{eqnarray}
L^{\varphi}_{h_rVV}&=& \frac{g^2_5}{4y_c}h_r V^\mu V_\mu \int{ \varphi(y) \phi_0(y)e^{-2ky} dy} \nonumber \\
&=&\frac{g^2_5}{4y_c}h_r V^\mu V_\mu \int{ \frac{1}{\Lambda_r e^{2ky_c}} \left( 2\sqrt{5} e^{4ky} \sqrt{\frac{1}{\kappa^2}}-\frac{13 e^{2 k y + 2 k y_c} \sqrt{\frac{1}{\kappa^2}} y}{\sqrt{5} y_c} \right) \ell \phi_0(y) e^{-ky} dy}  \nonumber \\
&=& -8\frac{m^2_V}{\Lambda_r}h_r V^\mu V_\mu ~,
\end{eqnarray}
so that its overall coupling to the gauge-bosons is
\begin{equation}
L_{h_rVV}=-9\frac{m^2_V}{\Lambda_r}h_r V^\mu V_\mu ~.
\end{equation}

The tree-level couplings of the Higgs-radion to the massless Gauge-bosons are the same as the couplings of the radion in the original GW scenario \cite{Radion-Xsec}, due to the fact that there is no
tree-level contribution from $\varphi$. In particular, these couplings arise from the scale anomaly, see e.g. \cite{Radion-Xsec}:
\begin{equation}
L_{h_r\gamma\gamma/h_rgg}=h_r \frac{1}{\Lambda_r
}\frac{\beta_{SU(3)/U(1)}}{2g}tr\left(F_{\mu\nu}F^{\mu\nu}\right) ~.
\end{equation}

\section{The 126 GeV Higgs-Radion phenomenology at the LHC}
The Higgs-radion decay widths can now be calculated from the couplings derived in the previous section. It is easy to verify that the difference (in the values of the various widths)
between our scenario and the pure radion scenario are some
numerical factors. In particular, using the results of \cite{Radion-Xsec}
for the pure radion case and incorporating the extra factors associated with the Higgs-radion couplings, we find:
\begin{eqnarray}
\Gamma(h_r \to WW^\star,ZZ^\star) &=& 81 \frac{v_{SM}^2}{\Lambda^2_r} \Gamma(h\to WW^\star,ZZ^\star)_{SM} ~, \\
\Gamma(h_r \to gg) &=& \frac{\alpha_s^2 m_h^3}{32\pi^3\Lambda^2_r}(b_{QCD}+4x_t(1+(1-x_t)f(x_t)))^2 ~,\\
\Gamma(h \to \gamma \gamma) &=& \frac{\alpha_{em}^2 m_h^3}{256\pi^3\Lambda^2_r}(b_2+b_y-9(2+3x_W+3x_W(2-x_W)f(x_W))+
\nonumber \\
&+&\frac{32}{3}x_t(1+(1-x_t)f(x_t)))^2 ~, \\
\Gamma(h_r \to b \bar b,c \bar c,\tau \bar\tau) &=& 81 \frac{v_{SM}^2}{\Lambda^2_r} \Gamma(h\to b \bar b,c \bar c,\tau \bar\tau)_{SM} ~,
\end{eqnarray}
where $x_{t},~x_{W} \equiv \textit{}\frac{4m_{t},~m_{W}^2}{m_{h_r}^2}$, $b_2=19/6$, $b_y=-41/6$, $b_{QCD}=7$ and $v_{SM}=246~GeV$. Also,
$f(z)$ is
\begin{equation}
f(z)=\left\{\begin{array}{c} (sin^{-1}(1/\sqrt{z}))^2 ~,  \quad z \ge 1 \\ -\frac{1}{4}\left(log \frac{1+\sqrt(1-z)}{1-\sqrt(1-z)}-i\pi \right) ~, \quad z<1 \end{array} \right.
\end{equation}

For the decays $h_r \to b \bar b$, $h_r \to c \bar c$  and $h_r \to \tau \bar\tau$ we have used the approximation that the wave functions of $\varphi$ and $\phi_0$ are much steeper than the wave functions of the $b$, $c$ and the $\tau$, so that the latter could be taken to be flat (see \cite{RS-flavor}).

For the 126 GeV data we use the data in Table 1 of \cite{global-Higgs-fit}, with the definition of $\chi^2$ thereof. Fitting $\Lambda_r$ to the data, we find that the 95\% CL allowed region for $\Lambda_r$ is $2.8~TeV<\Lambda_r<3.4~TeV$, where the best fitted value is $\Lambda_r=3.0$ TeV.
In particular, for $\Lambda_r=3.0~TeV$ the resulting values of the signal strengths in
the various channels are:
\begin{eqnarray}
\mu^{ggF}_{\gamma\gamma}(\Lambda_r=3.0~TeV)&=&1.45 \\
\mu^{VBF}_{\gamma\gamma}(\Lambda_r=3.0~TeV)&=&0.95\\
\mu^{ggF}_{VV}(\Lambda_r=3.0~TeV)&=&0.87\\
\mu^{VBF}_{VV}(\Lambda_r=3.0~TeV)&=& 0.57\\
\mu^{VH}_{bb}(\Lambda_r=3.0~TeV)&=& 0.57\\
\mu^{ggF}_{\tau\tau}(\Lambda_r=3.0~TeV)&=&0.87\\
\mu^{VBF}_{\tau\tau}(\Lambda_r=3.0~TeV)&=&0.57
\end{eqnarray}
where the superscripts denote the production mechanism and the subscripts denote the decay channel.
The agreement with the measured data is at the level of $1\sigma$,
 i.e., we obtain $\chi^2_{min} \approx 5$ for 5 dof. Note also that
our calculated signals scale with $\Lambda_r$ as
\begin{equation}
\mu(\Lambda_r) =\mu(\Lambda_r=3.0~TeV)\frac{(3.0~TeV)^2}{\Lambda^2_r} ~.
\end{equation}

The Higgs-radion branching ratios, which do not depend on $\Lambda_r$,
are compared to the SM Higgs ones in Table~\ref{tab1}.$^{[3]}$\footnotetext[3]{We thank Heather Logan for discussions
regarding Table \ref{tab1}.}

\begin{table}[htb]
\begin{center}
\begin{tabular}{|l||c|c|}
\hline
 & SM ($m_h=126~GeV$) & Higgs-Radion ($m_{h_r}=126~GeV$) \\
 \hline
$Br(h\to WW^*)$ & 0.231 &0.204 \\
$Br(h \to ZZ^*)$ & 0.0289 & 0.0257 \\
$Br(h \to gg)$ & 0.0848& 0.13\\
$Br(h \to \gamma\gamma)$ & $2.28 \cdot 10^{-3}$ & $3.8\cdot 10^{-3}$ \\
$Br(h \to b \bar b)$ & 0.561 &  0.545 \\
$Br(h \to \tau \bar\tau)$ & 0.0615 & 0.063\\
$Br(h \to c \bar c)$ & 0.0283 & 0.028 \\
Total width [GeV] & $4.21 \cdot 10^{-3}$ &  $2.2 \cdot 10^{-3}$ \\
\hline
\end{tabular}
\caption{\emph{The Higgs-radion and the SM Higgs branching ratios and total width. The SM
values are taken from \cite{Higgs-handbook}.}}
\label{tab1}
\end{center}
\end{table}

The KK scale, usually  defined by the mass of the lowest KK excitation of the gluon (the ``KK-gluon" \cite{Higgs-Radion-Mixing-2}),
is given in our case by (recall that we have $\frac{k}{M_{Pl}} \approx 1.6$):
\begin{equation}
M_{KKG}=\Lambda_r\frac{k}{M_{Pl}} \approx 1.6\Lambda_r \approx 4.8~TeV~ (for~\Lambda_r=3~TeV)
\end{equation}

Note that the GW radion interpretation of the 126 GeV data requires $\Lambda_r<1~TeV$ \cite{Radion125}, thus leading to a very small KK scale which is ruled out by the direct resonance searches at the LHC \cite{LHC-search}. Furthermore, a scenario with mixing between the pure radion and the Higgs state is allowed \cite{Higgs-Radion-Mixing,Higgs-Radion-Mixing-2}, but in order to be consistent with the data it requires the 126 GeV state to be Higgs-dominated. This is in clear contrast to our scenario where the Higgs-radion interpretation of the 126 GeV data sets $4.48~TeV<M_{KKG}< 5.44~TeV$, which is well above the exclusion from direct searches \cite{LHC-search}.

Before summarizing, we wish to add a few comments on some of the phenomenological implications of our findings.
\begin{itemize}
\item The list of branching ratios
above indicates that only $h_r \to gg$ and
$h_r \to \gamma \gamma$ are significantly
larger than in the SM, i.e., over 50\%.
We must stress, however, that in our considerations above, the 1-loop effect of the
KK tower of particles on the decay widths and the production mechanism of $h_r$ has not been taken into account; this may be especially relevant as
these modes ($h_r \to gg$ and
$h_r \to \gamma \gamma$) are known to be amenable
to sizeable corrections due to these effects ~\cite{Azatov_etal,
Carena_etal}.
While the photonic mode has already been the center of focus for
quite sometime, we want to emphasize the importance of a direct measurement of the glu-glu branching ratio.
This is clearly very challenging but attempts via $t \bar t h$ and  or via $t \bar b h$ may well
prove rewarding and deserve a high priority.
\item The allowed range of the KK-gluon masses
in our picture
is somewhat higher than what is expected to be accessible for direct production of KK-gluons
at the LHC with 14 TeV c.m. energy,  as was emphasized in ~\cite{Agashe_etal, Randall_etal}.
\item There are few other features of our coupled Higgs-radion
picture and its confrontation
with the LHC data that deserve brief remarks. First, since the KK-gluon mass needs to be around 5 TeV,
it may well be that custodial symmetry is needed to respect electroweak precision
constraints (EWPC)~\cite{custod-symm}. If so, our analysis can be extended to include a custodial symmetry in the bulk.
 Furthermore, one of the most compelling rationale for the RS-warped idea is that it can be invoked to quite naturally understand flavor hierarchies \cite{RS-flavor}. However, in its simplest (``anarchic") implementation (see Agashe {\it et al} in~\cite{RS-flavor}),  the resulting FCNCs acquire some right-handed components which have enhanced effect on Kaon mixing ~\cite{Kaon-mixing}.
 It, therefore, becomes difficult to lower the KK-gluon mass below $\sim 10$ TeV without enforcing some additional flavor symmetry \cite{Perez-Randall}
and/or some mild tuning \cite{Monika_etal, Matthias_etal}. We hope to return to address these issues pertaining to EWPC and flavor in conjunction with our Higgs-radion idea in the near future.
\item It is important to note that our solution suffers
from a ``little-hierarchy" problem, i.e. $\frac{v^2_{eff}}{M^2_{KKG}}\approx 2.5\cdot 10^{-3}$.
It will be interesting to see whether the solution to this problem in the RS scenario can be implemented here as well. The most notable idea is that of the Higgs being a pseudo Nambu-Goldstone boson \cite{MCHM}, living in a coset of SO(5)/SO(4). This is manifested on the AdS side as the fifth component of the SO(5)/SO(4) gauge boson in the bulk \cite{Hosotani}, the effective potential for which is generated due to radiative corrections. This is in fact some form of a bulk Higgs with a VEV profile, and it is conceivable that a stabilization of the radius could be achieved here as well.
\end{itemize}

\section{summary}
We have explored a scenario where the EW-Planck hierarchy is stabilized by the presence of an $SU(2)_L$ scalar doublet in the bulk of an RS setup. In particular, we have showed that the 5d VEV profile of this bulk doublet can have a dual role: breaking the EW symmetry and at the same time stabilizing the radius of the warped extra dimension, i.e, the EW-Planck hierarchy of scales.

The conditions for the stabilization were derived and found to be different from the conventional RS scenario which employ the Goldberger-Wise (GW) mechanism. Our setup does not require any additional fine-tuning other than the one associated with the 4d cosmological constant, also present in the GW framework.

The lowest excitation of the coupled SU(2) scalar doublet--gravity system,
denoted here as the ``Higgs-radion",
is identified with the recently discovered 126 GeV scalar.
We derive the couplings of our Higgs-radion which turn out to be
significantly different from those of a pure radion in RS models.
This is due to the fact
that the Higgs-radion state has a component in
the 5d bulk doublet that has direct couplings to the SM fields.

An obvious but important consequence of our model is that it implies that a separate
low-lying scalar (radion) of the conventional KK setup does not exist.

We calculate the Higgs-radion production and decay processes relevant for the LHC, and find that it is compatible with all the present Higgs data for a KK gluon mass of the order of 5 TeV. Since the Higgs-radion signals are
found to deviate from the SM in some of the channels (e.g.,
$h_r \to \gamma \gamma$ and $h_r \to gg$), we expect that with more data
it will be possible to verify or exclude our setup.

\bigskip

{\bf Acknowledgments:}
We are extremely grateful to Kaustubh Agashe for many very valuable discussions.
We also benefitted from useful conversations with Hooman Davoudiasl, Yael Shadmi, Yotam Soreq,
John Terning and Tomer Volansky.  SBS and MG acknowledge research support from the Technion and the work of AS is supported in part
by the US DOE contract No. DE-AC02-98CH10886 (BNL).

\end{document}